\documentclass[submission, Proceedings]{SciPost}
 \usepackage{delimset} 

\newcommand{\dd}{\mathrm{d}}
\newcommand{\alg}[1]{\mathfrak{#1}}
\newcommand{\grp}[1]{\mathrm{#1}}
\newcommand{\gen}[1]{\mathrm{#1}}
\newcommand{\comm}[2]{[#1,#2]}
\newcommand{\order}[1]{\mathcal{O}(#1)}
\newcommand{\superN}{\mathcal{N}}

\newcommand{\tr}{\mathrm{Tr}}

\newcommand{\mbf}[1]{{#1}}

\newcommand{\gf}[1]{\mathcal{#1}}

\ifx\genfrac\sdflkaj
\newcommand{\atopfrac}[2]{{{#1}\above0pt{#2}}}
\else
\newcommand{\atopfrac}[2]{\genfrac{}{}{0pt}{}{#1}{#2}}
\fi

\definecolor{darkred}{RGB}{180,30,30}

\newcommand{\highlight}[1]{\textcolor{darkred}{#1}}
\newcommand{\sfrac}[2]{{\textstyle\frac{#1}{#2}}}
\newcommand{\half}{\sfrac{1}{2}}

\makeatletter
\newlength{\apb@width}
\newcommand{\autoparbox}[2][c]{\settowidth{\apb@width}{#2}\parbox[#1]{\apb@width}{#2}}
\newcommand{\includegraphicsbox}[2][]{\autoparbox{\includegraphics[#1]{#2}}}
\makeatother

\makeatletter
\def\mr@ignsp#1 {\ifx\:#1\@empty\else #1\expandafter\mr@ignsp\fi}%
\newcommand{\multiref}[1]{\begingroup
\xdef\mr@no@sparg{\expandafter\mr@ignsp#1 \: }%
\def\mr@comma{}%
\@for\mr@refs:=\mr@no@sparg\do{\mr@comma\def\mr@comma{,}\ref{\mr@refs}}%
\endgroup}
\makeatother

\newcommand{\hypref}[2]{\ifx\href\asklfhas #2\else\href{#1}{#2}\fi}
\newcommand{\Secref}[1]{Section~\multiref{#1}}
\newcommand{\Figref}[1]{Figure~\multiref{#1}}

\hypersetup{
    colorlinks,
    linkcolor={red!50!black},
    citecolor={blue!50!black},
    urlcolor={blue!80!black}
}

\usepackage[bitstream-charter]{mathdesign}
\DeclareSymbolFont{usualmathcal}{OMS}{cmsy}{m}{n}
\DeclareSymbolFontAlphabet{\mathcal}{usualmathcal}

\begin{document}


\begin{center}{\Large \textbf{
Integrability for Feynman Integrals\\
}}\end{center}

\begin{center}
Florian Loebbert
\end{center}

\begin{center}
Bethe Center for Theoretical Physics, University of Bonn 
\\
Nussallee 12, 53115 Bonn, Germany
\\[2pt]
loebbert@uni-bonn.de
\end{center}

\begin{center}
\today
\end{center}


\definecolor{palegray}{gray}{0.95}
\begin{center}
\colorbox{palegray}{
  \begin{tabular}{rr}
  \begin{minipage}{0.1\textwidth}
    \includegraphics[width=20mm]{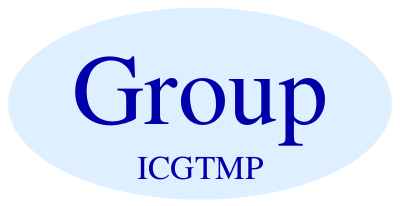}
  \end{minipage}
  &
  \begin{minipage}{0.85\textwidth}
    \begin{center}
    {\it 34th International Colloquium on Group Theoretical Methods in Physics}\\
    {\it Strasbourg, 18-22 July 2022} \\
    \end{center}
  \end{minipage}
\end{tabular}
}
\end{center}

\section*{Abstract}
{
We give a brief overview of the Yangian symmetry of Feynman integrals. After a short introduction to the Yangian and integrability, we motivate the emergence of integrable structures for Feynman integrals via the fishnet limit of AdS/CFT. We discuss the resulting Yangian differential equations for massless fishnets in four dimensions as well as generalizations to massive propagators and generic dimensions. We also comment on the relation to momentum space conformal symmetry and on examples in dimensional regularization. Finally we sketch the recent application to fishnet integrals in two spacetime dimensions and the curious identification of Yangian invariants with period integrals of Calabi--Yau geometries.
}

\vspace{10pt}
\noindent\rule{\textwidth}{1pt}
\tableofcontents\thispagestyle{fancy}
\noindent\rule{\textwidth}{1pt}
\vspace{10pt}

\newpage
\section{Introduction}
\label{sec:intro}
Integrable models appear in all areas of physics, from classical mechanics to quantum field theory. While they have a natural home in two-dimensional systems like spin chains and 2d field theories, applications in four-dimensional particle physics have long been limited to special high-energy limits in QCD \cite{Lipatov:1993yb}. From the integrability viewpoint an important (still formal) step into the direction of particle phenomenology was the discovery of integrable structures in a four-dimensional quantum field theory, i.e.\ planar $\mathcal{N}=4$ super Yang--Mills (SYM) theory \cite{Beisert:2010jr}. While also here a clear connection to two-dimensional physics in the AdS/CFT-dual string theory is present, this finding opened a new door to apply and extend the toolbox of integrability. Here we discuss another step into this direction, namely the appearance of integrable structures for Feynman integrals, which constitute the building blocks of generic quantum field theories including those realized in nature. On the one hand, this detaches integrability in four dimensions from the special nature of $\mathcal{N}=4$ SYM theory. On the other hand, an explanation for these mathematical structures of Feynman integrals can still be found in the integrability of the planar AdS/CFT correspondence via its so-called fishnet limits \cite{Gurdogan:2015csr}.
 The simplified fishnet theories arise as particular double-scaling limits of the so-called gamma-deformed $\superN=4$ SYM theory and they have the particular feature that their correlation functions are in one-to-one correspondence with individual Feynman integrals \cite{Caetano:2016ydc,Chicherin:2017cns}. Notably, some of the integrability structures which underly the planar AdS/CFT duality are inherited by these elementary Feynman integrals as discussed below. Here we will focus on the so-called Yangian symmetry of these integrals, which is an infinite dimensional extension of a Lie algebra and can be understood as the algebraic foundation of certain classes of integrable models.
The need to study the mathematical properties and symmetry structures of Feynman integrals is underlined by the fact that their computation still represents a bottle neck for phenomenological predictions, see \cite{Bourjaily:2022bwx} for a recent review. At lower loop orders they are governed by the class of multiple polylogarithms which are defined as iterated integrals with rational integration kernels.
In more general cases elliptic integrals and worse geometric structures (e.g.\ Calabi--Yau geometries) appear. These are typically characterized by roots of higher order polynomials in the denominator of the integrand.
The exploration of multi-loop Feynman integrals and their mathematical structure is currently a very active field of research, which makes the appearence of integrable structures even more fascinating.

\section{Integrability and the Yangian}

As indicated above, integrability is rooted in two-dimensional physics where it is often identified with the concept of \emph{factorized scattering}. The latter denotes the phenomenon that the $n$-body scattering matrix of a given theory factorizes into two-body scattering events.%
\footnote{Of course, this is only possible in models which feature a notion of scattering.} 
In fact, the idea of factorized scattering may be considered the closest to a proper definition of quantum integrability which to date is still lacking, see e.g.\ \cite{sutherland2004beautiful, Caux:2010by}.
\begin{figure}[t]
\begin{center}
\includegraphicsbox[scale=.8]{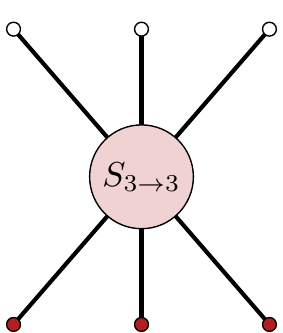}
$=$\includegraphicsbox[scale=.8]{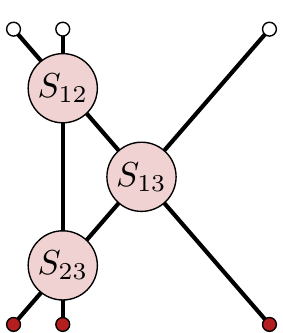}
$=$
\includegraphicsbox[scale=.8]{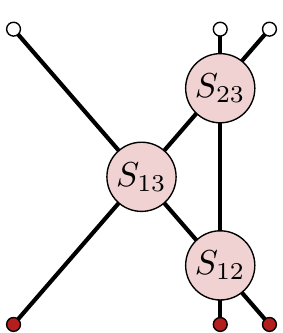}
\caption{Factorized scattering in two dimensions. For consistency the quantum Yang--Baxter equation (right equality) has to hold: $\gen{S}_{12}S_{13}S_{23}=S_{23}S_{13}S_{12}$. A similar equation holds for the generating function $T(u)$ of the Yangian generators together with the quantum R-matrix: $RTT=TTR.$}
\label{fig:FacScat}
\end{center}
\end{figure}
On the one hand, the implications of factorized scattering include the applicability of powerful solution techniques such as the celebrated Bethe Ansatz. On the other hand, the reason for factorized scattering  is found in the existence of a tower of conserved charges or higher symmetries. In spacetime dimensions greater than two, higher symmetries are believed to imply a trivial S-matrix via the Coleman--Mandula theorem \cite{Coleman:1967ad}. Two-dimensional models, however, provide a fascinating loop-hole and allow for non-trivial scattering and higher symmetries at the same time. In fact, one can make the connection between these higher symmetries and the factorization of the S-matrix more precise and identify a set of symmetry generators $\gen{\widehat J}$ such that $\comm{\gen{\widehat J}}{S}=0$ implies the factorization of $S$, cf.\ \cite{Loebbert:2018lxk}. 
The notion of factorized scattering reduces the integrable S-matrix in two dimensions to the two-body scattering matrix as its fundamental building block, cf.\ \Figref{fig:FacScat}.  In the simplest situations, this two-body matrix is a rational function of the rapidity parameter. This case of rational quantum integrable models is related to the so-called Yangian, an infinite dimensional extension of a Lie algebra $\alg{g}$ that was introduced by Drinfeld in 1985 \cite{Drinfeld:1985rx}, see also the reviews \cite{Bernard:1992ya,MacKay:2004tc,Loebbert:2016cdm}. The Yangian $Y[\alg{g}]$ in its so-called \emph{first realization} is defined by two sets of generators which are characterized by their tensor product actions of the following form:%
\footnote{In the following these tensor products will be identified with the external legs of planar Feynman graphs.}
\begin{align}
&\text{Level 0}:&  &\gen{J}^a=\sum_{k=1}^n \gen{J}^a_k ,
\\[-1mm]
&\text{Level 1}: & &
\gen{\widehat J}^a=f^a{}_{bc} \sum_{j<k=1}^n \gen{J}_j^c \,\gen{J}_k^b+\sum_{k=1}^n  s_k\gen{J}^a_k,
\label{eq:deflev1}
\\[2pt]
&\text{Serre relations:}&&
\comm{\gen{\widehat J}_a}{\comm{\gen{\widehat J}_b}{\gen{ J}_c}}
-
\comm{\gen{J}_a}{\comm{\gen{\widehat J}_b}{\gen{\widehat  J}_c}}=\order{\gen{J}^3}.
\end{align}
Here the local level-0 operators generate the underlying Lie algebra and have a trivial coproduct.
 Their densities $\gen{J}^a_k$ are also employed to construct the bilocal level-one generators $\gen{\widehat J}_a$ as specified above. All higher generators of this infinite dimensional algebra can be obtained by iterative commutation. Here the Serre relations, a quantum algebra generalization of the Lie algebra's Jacobi identity, provide additional constraints on the representation.
Note that the Serre relations for the case of the differential operator representation of the conformal algebra, which becomes relevant below, have been discussed in \cite{Miczajka:2022jjv,Dokmetzoglou:2022mfd}.
 The so-called evaluation parameters $s_k$ in the above definition parametrize an external automorphism of the Yangian algebra which is realized by the Lorentz boost symmetry in relativistic models. 

The Yangian has been studied in the context of various physical setups; examples include
the Heisenberg spin chain with $\alg{g}=\alg{su}(2)$, cf.\ \cite{Loebbert:2016cdm};
the AdS/CFT duality with  $\alg{g}=\alg{psu}(2,2|4)$, cf.\ \cite{Torrielli:2010kq};
or  Euclidean fishnet integrals in four dimensions with $\alg{g}=\alg{so}(1,5)$, cf.\ \cite{Chicherin:2017cns}.

 In fact, the above Yang--Baxter equation depicted in \Figref{fig:FacScat} can be lifted to a purely algebraic structure relating the so-called quantum R-matrix and the monodromy matrix $T$ via the RTT-relations: 
  \begin{equation}
 R_{12}(u-v)T_1(u)T_2(v)=T_2(v)T_1(u)R_{21}(u-v).
 \end{equation}
 Here the commutation of two monodromy matrices $T(u)$ acting on spaces 1 and 2, is encoded in the Yangian R-matrix. The expansion of $T(u)$ in the spectral parameter $u$ gives rise to the different levels of the Yangian generators. This so-called \emph{RTT-realization} of the Yangian represents an alternative way to define the same algebra, which is closely related to the concept of factorized scattering in physical models.

\section{AdS/CFT and Fishnet Theories}

One of the physical setups where the Yangian algebra plays a crucial role is the planar AdS/CFT correspondence.
Here both sides of the duality, i.e.\ $\superN=4$ SYM theory and IIB string theory on $\mathrm{AdS}_5\times \mathrm{S}^5$ feature the Lie algebra symmetry $\alg{psu}(2,2|4)$. In the planar limit, this symmetry extends to the Yangian $Y[\alg{psu}(2,2|4)]$ with a natural supersymmetric generalization of the above algebra definitions. The Yangian has been investigated in many different setups on the gauge and string side of the duality. It first appeared in this context as a symmetry of the dilatation generator of $\superN=4$ SYM theory \cite{Dolan:2003uh}. Here a consequence of Yangian symmetry is that the spectrum of the dilatation operator can be computed via appropriate generalizations of the Bethe Ansatz technique that was originally designed for the $\alg{su}(2)$ Heisenberg spin chain~\cite{Beisert:2010jr}. Notably, $\mathcal{N}=4$ SYM theory represents the first four-dimensional quantum field theory which is believed to be completely integrable%
\footnote{A precise definition of this statement is hard to provide.} 
 in the planar limit. It is defined by the following schematic Lagrangian with $\grp{SU}(N)$ gauge symmetry:
\begin{align}
\mathcal{L}_{\superN=4}
\simeq \,\mathrm{Tr} \big(&
  -\gf{F}\gf{F}
- \gf{D} \Phi \gf{D} \Phi 
+ \bar\Psi\gf{D}\Psi
- g^2 \comm{\Phi}{\Phi}^2
-  g   \Psi\comm{\Phi}{\Psi}
-  g \bar\Psi\comm{\Phi}{\bar \Psi}
\Big).
\label{eq:LagrN4}
\end{align}
It includes six  matrix-valued scalars $\Phi_m$, four Dirac spinors $\Psi_A$ and $\bar \Psi^A$ and the gauge field strength $\mathcal{F}$. 
Remarkably, the $\beta$-function of $\superN=4$ SYM theory is zero which makes the theory quantum conformal. Integrability arises in the large-$N$ limit with fixed t'Hooft coupling $\lambda=g^2 N$, where non-planar Feynman diagrams are suppressed by factors of $1/N$.

In the above model an additional set of three parameters $\gamma_1$, $\gamma_2$ and $\gamma_3$ can be introduced via the so-called gamma-deformation of $\superN=4$ SYM theory. All products of fields in the Lagrangian $\mathcal{L}_{\superN=4}$ are replaced by non-commutative products, which leads to phase factors  $e^{i\gamma_j (\dots)}$ in front of the different terms in the resulting $\mathcal{L}_{\mathcal{N}=4}^\gamma$. Here the $(\dots)$ in the phase factor depends on the different $\grp{SU}(4)$ Cartan charges of the fields. The additional parameters in the gamma-deformed Lagrangian now allow for interesting double-scaling limits as noted in~\cite{Gurdogan:2015csr}. After rescaling the fields by $\sqrt{N},$ one takes $\lambda\to 0$ and sends the gamma-parameters to imaginary infinity, i.e.\ $\gamma_j \to i\infty$, while keeping the new couplings $\xi_j= \sqrt{\lambda} e^{-i\gamma_j/2}$ constant.
In the simplest case only one of the couplings, e.g.\ $\xi=\xi_3$, is non-vanishing and most of the fields decouple. The result is the following Lagrangian of the bi-scalar fishnet model for two complex matrix-valued fields denoted by $X$ and $Z$:%
\footnote{
We note that in general this bi-scalar Lagrangian is not complete and requires additional double-trace terms for conformality \cite{Sieg:2016vap,Grabner:2017pgm}. For the Feynman graphs (alias correlators) considered in this paper,  these double-trace terms do not play a role.}
\begin{equation}
\mathcal{L}_\text{F}=
N\, \mathrm{Tr}\brk*{
-\partial_\mu \bar X \partial^\mu X 
-
\partial_\mu \bar Z \partial^\mu Z
+\xi^2 \bar X\bar Z X Z
}.
\label{eq:LagrFish}
\end{equation}
One of the remarkable properties of this model is that a large class of its planar correlation functions is computed by single Feynman integrals of fishnet structure. The correlator $\langle XXXZZ\bar XZ\bar X \bar X\bar X\bar Z\bar Z X \bar Z \rangle$ for instance corresponds to the following position space Feynman graph:
\begin{equation}
\includegraphicsbox{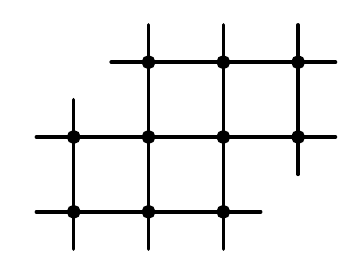}
\end{equation}
Here blobs represent integration vertices for the spacetime coordinates $x^\mu$ with $\mu=0,\dots,3$ and lines stand for propagators as summarized in the following Feynman rules:
\begin{equation}
\includegraphicsbox{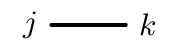}
\to\frac{1}{x_{jk}^2},
\qquad\qquad
\includegraphicsbox{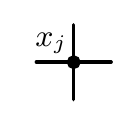}
\to 
\int \dd^4 x_j.
\end{equation}
Note that we write $x_{jk}^\mu=x_j^\mu-x_k^\mu$ for differences of the Euclidean spacetime vectors and we omit the color structure of the graph.
Any graph that can be cut out of a square fishnet lattice thus represents a correlator in the above fishnet theory. This drastic limitation to a single contributing Feynman graph is attributed to the chiral four-point vertex in the fishnet Lagrangian.%
\footnote{This chirality is also responsible for the model being non-unitary.}
The fact that the four-point vertex in \eqref{eq:LagrFish} is chiral implies a particular admissible orientation of the color flow in the diagram.

While the fishnet model has been investigated from various perspectives, in the following we will focus on a particular implication, namely that Feynman integrals possess a higher Yangian symmetry.
\section{Integrability for Feynman Integrals}
\label{sec:intforF}

A conformal Lie algebra (or Yangian level-0) symmetry of the above Euclidean Feynman integrals is realized via the following differential operator representation of $\alg{so}(1,5)$:
\begin{align}
\gen{J}^a =\sum_{j=1}^n \gen{J}_j^a
\qquad \text{with}\qquad
\gen{J}_j^a \in
\begin{cases}
&\gen{D}_j=-i x_{j\mu} \partial_{x_j}^{\mu}-i \Delta_j,\\
&\gen{L}_j^{\mu\nu}=ix_j^\mu \partial_{x_j}^\nu-i x_j^\nu \partial_{x_j}^\mu, \\
&\gen{P}_j^\mu=-i\partial_{x_j}^\mu,\\
&\gen{K}_j^\mu=ix_j^2 \partial_{x_j}^\mu-2 i x_j^\mu x_{j\nu} \partial_{x_j}^{\nu} -2 i \Delta_j x_j^\mu.
\end{cases}
\end{align}
Here derivatives act on the external legs of the Feynman graph and the scaling dimensions $\Delta_j$ are set to 1 for the scalar particles considered.
Invariance under the above conformal generators implies that an $n$-point fishnet integral can be written as a product of a prefactor $V_n$,  which carries the conformal weight of the integral, and a conformal function that only depends on a set of conformal variables defined in terms of cross ratios:
\begin{equation}
I_n=V_n\, \phi(u_1,u_2,\dots).
\label{eq:defconffunc}
\end{equation}
Here the precise number of cross ratios $u_k$ depends on the number of external points $n$ of the Feynman graph.

The level-1 Yangian symmetry on the other hand can be constructed from the Lie algebra generator densities given above with the prescription dictated by \eqref{eq:deflev1}. 
Given full level-0 invariance of a Feynman integral $I_n$, the Yangian commutation relations imply that invariance under a single level-1 generator yields full Yangian symmetry of the graph. Here one typically works with the level-1 momentum generator whose explicit form is given by
\begin{equation}
\gen{\widehat P}^{\mu}
=\sfrac{i}{2} \sum_{j,k=1}^n 
\text{sign}(k-j)\brk!{\gen{P}_j^{\mu} \gen{D}_k + \gen{P}_{j\nu} \gen{L}_k^{\mu \nu} } + \sum_{j=1}^n s_j \gen{P}_j^{ \mu}.
\end{equation}

The invariance of the above fishnet integrals under the Yangian can be proven with the so-called lasso method and the help of the monodromy matrix $T(u)$, whose expansion coefficients around $u=\infty$ are essentially the Yangian generators, cf.\ \cite{Chicherin:2017cns}. At order $k$ of the $1/u$ expansion these correspond to $k$-local generators, while the whole monodromy $T(u)$ acts on all $n$ legs of a given graph and is constructed as a product of $n$ Lax operators. Schematically, this action of $T(u)$ can be depicted by a line encircling the Feynman diagram similar to a lasso. A set of commutation relations and identities for the Lax operators and propagators allow to disentagle the monodromy from this graph, which results in an eigenvalue equation depicted by (see \cite{Chicherin:2017cns} for details)
\begin{equation}
\includegraphicsbox{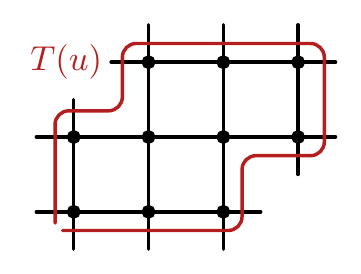}
=
f(u)\includegraphicsbox{FigFishnetGraph.pdf}
\end{equation}
Here $f(u)$ denotes a polynomial eigenvalue function of the spectral parameter.
Remembering that $T(u)\simeq 1+1/u\, \gen{J}+1/u^2 \,\gen{\widehat J}+\dots$, this eigenvalue equation implies the level-0 and level-1 invariance of the given integral for a judicious choice of the evaluation parameters~$s_k$, see the prescription in \cite{Loebbert:2020hxk}.

While level-0 conformal invariance of a given Feynman graph implies the form \eqref{eq:defconffunc}, the level-1 symmetry yields additional constraints for the function $\phi$ of the conformal cross ratios. The invariance under the level-1 momentum generator can be rewritten in the following form~\cite{Loebbert:2019vcj}:
\begin{equation}
\qquad\quad 0 = \gen{\widehat P}^{\mu}  I_n = V_n \sum_{j<k=1}^n \frac{x_{jk}^\mu}{x_{jk}^2} \, D_{jk} \phi.
\end{equation}
Here at least for lower numbers of external points one can argue that the vectors ${x_{jk}^\mu}/{x_{jk}^2}$ are in fact independent, which implies that the Yangian invariance can be translated into 
a system of partial differential equations (PDEs) in the cross ratios:
\begin{align}
D_{jk}\, \phi = 0,
 \qquad 1\leq j<k\leq n .
 \label{eq:pdesys}
\end{align}
These coupled differential equations are highly constraining and their solutions provide a set of building blocks, whose linear combination represents the Feynman integral.

\paragraph{Example: 4D Cross (or Box) Integral.}

The simplest four-dimensional example is the Euclidean cross (or box) integral, whose two names refer to the two alternative representations in $x$- or $p$-space:
\begin{equation}
I_4^{4D}=
\includegraphicsbox[scale=1.2]{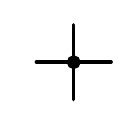}
=\int  \frac{\dd^4 x_0}{x_{10}^2x_{20}^2x_{30}^2x_{40}^2}
=\int \frac{\dd^4 \ell}
{\ell^2 (\ell+p_1)^2 (\ell+p_1+p_2)^2 (\ell-p_4)^2}.
\end{equation}
Here the coordinates $x_j^\mu$ can alternatively be interpreted as positions, or dual momenta via the relation to the ordinary momenta $p_j^\mu=x_j^\mu-x_{j+1}^\mu$.
The level-0 Yangian, or (dual) conformal symmetry, of the four-point integral $I_4$ implies the form
$I_4= \frac{1}{x_{13}^2 x_{24}^2} \phi(z,\bar z)$, 
with the conformal variables $z$ and $\bar z$ defined via the following relation to the cross ratios $u$ and $v$:
\begin{equation}
z\bar z=\frac{x_{12}^2 x_{34}^2}{x_{13}^2 x_{24}^2}=u, 
\qquad\qquad
(1-z)(1-\bar z)=\frac{x_{14}^2 x_{23}^2}{x_{13}^2 x_{24}^2}=v. 
\end{equation}
The additional
Yangian level-1 symmetry yields two coupled differential equations of the form (cf.\ \eqref{eq:pdesys} and  \cite{Loebbert:2019vcj})
\begin{align}
[D_j(z)-D_j(\bar z)]\phi(z,\bar z)=0,
\qquad
j=1,2,
&
\end{align}
with the second order differential operators $D_j$ given by
\begin{align}
&D_1(z)=
z (z-1)^2 \partial_z^2 +(3 z-1) (z-1)
  \partial_z+z,
\\
&D_2(z)=
z^2(z-1)  \partial_z^2 +(3 z-2) z
   \partial_z+z.
\end{align}
One finds four solutions to these equations which are of the form $f_j(z,\bar z)/(z-\bar z)$ with \cite{Loebbert:2019vcj}
\begin{align}
f_1&=1,\nonumber
\\
f_2&=\log(\bar z)-\log (z ),
\label{eq:fourinvariants}
\\
f_3&=\log(1-\bar z) -\log (1- z),\nonumber
\\
f_4&=2 \text{Li}_2(z) - 2 \text{Li}_2(\bar z) 
		+ \log \frac{1-z}{1-\bar z} \log (\bar z z) .\nonumber
\end{align}
In addition to the Yangian symmetry the cross integral is invariant under all permutations of its external legs.
This permutation invariance singles out the Bloch--Wigner dilogarithm $f_4$  as the correct Yangian invariant which represents this integral. While this solution has been known since the early works \cite{Usyukina:1992jd}, it is remarkable that the cross can be bootstrapped using integrability and permutation symmetry only \cite{Loebbert:2019vcj}. We note that the overall constant prefactor remains to be fixed by other means, e.g.\ numerics or a coincidence limit of external points.

While one might expect that integrability fixes an observable completely, here we have employed the additional permutation symmetry to single out the correct Yangian invariant of the four solutions \eqref{eq:fourinvariants}. It turns out that all four building blocks are required to express the integral when going from Euclidean to Minkowski kinematics \cite{Corcoran:2020akn}. As opposed to a single Euclidean region, in the Minkowski case one distinguishes 64 kinematic regions depending on the signs of the $x_{jk}^2$.  Here, in a given region the integral is typically invariant under a subset of all permutations and $f_4$ is no longer the only admissible Yangian invariant. Via similar reasoning as above one can bootstrap the Minkowski integral using its Yangian symmetry modulo a small number of constant coefficients \cite{Corcoran:2020epz}.

\section{Generalizations: Dimensions, Masses,  Momentum Space,\dots }

In this section we indicate some generalizations, which go beyond massless square fishnet integrals in four spacetime dimensions.


\paragraph{Fishnet Structures and Dimensions.}

The above bi-scalar fishnet model represents the simplest double-scaling limit of gamma-deformed $\superN=4$ SYM theory. More involved models are obtained by combining different limits of the three parameters $\gamma_j$ and the coupling constant. In particular, models exist where also some of the fermions survive, which results e.g.\ in Yangian-invariant Feynman graphs of brick wall structure \cite{Chicherin:2017frs}.
Here the non-scalar particles require non-scalar representations of the conformal algebra to construct the Yangian generators.

Apart from generalizing the field content, the above fishnet theories have natural generalizations to different spacetime dimensions. In particular, there exists a  double-scaling limit of Aharony--Bergman--Jafferis--Maldacena (ABJM) theory which results in a three-dimensional fishnet model with triangular Feynman graphs~\cite{Caetano:2016ydc}. Also in six spacetime dimensions one can write down a similar fishnet theory with hexagonal graph structure~\cite{Mamroud:2017uyz}.
All of the associated Feynman graphs feature the above  Yangian symmetry over the conformal algebra in the respective spacetime dimension \cite{Chicherin:2017frs}. This even generalizes to graphs with deformed propagator powers $a_j$ as long as these sum up to the spacetime dimension at each integration vertex \cite{Chicherin:2017frs,Loebbert:2019vcj,Loebbert:2020hxk}:
\begin{equation}
\frac{1}{x_{jk}^2} \quad \to \quad \frac{1}{(x_{jk}^{2})^{a_j}},
\qquad\qquad
\sum_{j\in\text{vertex}} a_j=D.
\label{eq:propandcond}
\end{equation}
If the latter conformal condition does not hold, one still finds an invariance under the level-one momentum generator, which relates to the below discussion of momentum space conformal symmetry.
Finally, the square fishnet model was generalized to a $D$-dimensional Lagrangian of the form \cite{Kazakov:2018qbr}
\begin{equation}
\mathcal{L}_\text{F}^D= N\,\tr \brk[s]*{ X(-\partial_\mu \partial^{\mu})^{\frac{D}{4}} \bar{X}+ Z (-\partial_\mu\partial^{\mu})^{\frac{D}{4}} \bar{Z}+\xi^2 XZ\bar{X}\bar{Z}}.
\label{eq:DFishLagr}
\end{equation}
Here the operators $(\partial_\mu \partial^{\mu})^{\frac{D}{4}}$ are understood as integral operators for the case $D\neq 4$.
The fact that integrability properties are preserved by all the above modifications demonstrates the universality of these mathematical structures. 

\paragraph{Yangian Symmetry for the Masses.}

In the massless case, we argued that Feynman integrals can be interpreted as correlators in the fishnet theory, which is connected to the integrable $\mathcal{N}=4$ SYM theory by means of the above double scaling limit. It is well known how to introduce masses in the latter theory via the Higgs mechanism, but no integrable structures are known that survive this process. Hence, a priori one might not expect to find integrable structures in massive Feynman integrals following the same logic. Still it is instructive to look at the massive version of $\superN=4$ SYM theory on the so-called Coulomb branch.
Masses are introduce by giving a vacuum expectation value (VEV) to one of the scalar fields in the Lagrangian \eqref{eq:LagrN4} \cite{Alday:2009zm}:
\begin{equation}
\hat \Phi=\langle \Phi\rangle +\Phi
\hspace{2cm}
\includegraphicsbox[scale=.25]{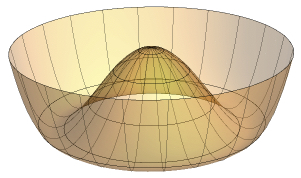}
\end{equation}
This leads to the so-called Coulomb phase of the theory and implies the appearence of massive propagators of the form
\begin{equation}
\frac{1}{\hat x_{jk}^2}= \frac{1}{{x_{jk}^2+(m_j-m_k)^2}}.
\end{equation}
Importantly, here the masses enter in differences in analogy to the spacetime coordinates.
In particular, for a judicious choice of the above VEV only planar Feynman integrals with masses on the boundaries survive the large-$N$ limit \cite{Alday:2009zm,Loebbert:2020tje}. These have been shown to posess a massive extension of the dual conformal symmetry with generators of the following form \cite{Alday:2009zm}:
\begin{align}
\gen{J}^a =\sum_{j=1}^n \gen{J}_j^a
\qquad \text{with}\qquad
\gen{J}_j^a \in
\begin{cases}
&\gen{D}_j = -i \brk!{x_{j\mu} \partial_{x_j}^\mu + \textcolor{darkred}{m_j \partial_{m_j}} + \Delta_j},\\
&\gen{L}_j^{\mu\nu} = i x_j^{\mu} \partial_{x_{j}}^{\nu} - ix^{\nu}_j \partial_{x_{j}}^{\mu}, \\
&\gen{P}^{\mu}_j = -i \partial_{x_{j}}^{\mu},\\
& \gen{K}^{\mu}_j = -2ix_j^{\mu}\brk!{x_{j\nu}  \partial_{x_j}^\nu  +  \textcolor{darkred}{m_j\partial_{m_j}} + \Delta_j} +i (x^2_j + \textcolor{darkred}{m^2_j})\partial_{x_{j}}^{\mu}.
\end{cases}
\end{align}
Here the mass can be interpreted as the $(D+1)$th component of the spacetime vector: $x_j^{D}=m_j$.
Unfortunately to date no massive Yangian symmetry has been detected in the full Coulomb phase of $\superN=4$ SYM theory. Still we can employ the definition of the level-1 generators in terms of the Lie algebra generator densities as given in \eqref{eq:deflev1}. Applying these to Feynman integrals with massive propagators, it turns out that planar one- and two-loop integrals in generic kinematics are annihilated for an appropriate choice of evaluation parameters \cite{Loebbert:2020hxk}. This statement can be proven explicitly for generic number of external legs and propagator powers obeying the conformal condition $\sum_{j\in\text{vertex}} a_j=D$ at each integration vertex \cite{Loebbert:2020glj}. The only criterion for this symmetry is that the  internal loop-to-loop propagator of the two-loop diagrams remains massless. Numerical tests at higher loop orders suggest that in fact all Feynman graphs cut out of a regular tiling of the plain feature this massive Yangian symmetry as long as all internal propagators remain massless \cite{Loebbert:2020glj}. Also a massive extension of the fishnet Lagrangian can be defined via a double-scaling limit of a gamma-deformed version of massive Coulomb branch $\superN=4$ SYM theory \cite{Loebbert:2020tje}:
\begin{align}
\mathcal{L}_\text{MF}=
&N\, \tr\brk*{
-\partial_\mu \bar X \partial^\mu X 
-
\partial_\mu \bar Z \partial^\mu Z
+\xi^2 \bar X\bar Z X Z
}
\nonumber\\
&
-  N(m_a - m_b)^2 {X^a}_b{\bar X^b}_a
- N(m_a - m_b)^2 {Z^a}_b{\bar Z^b}_a.
\end{align} 
The planar off-shell amplitudes in this theory correspond to the massive versions of Yangian invariant Feynman integrals described above.%
\footnote{Alternatively one can introduce masses via spontaneous symmetry breaking in the bi-scalar fishnet theory which, however, leads to different propagators and seems not to allow for integrable symmetries~\cite{Karananas:2019fox}.}


\paragraph{Momentum Space Conformal Symmetry.}

Notably, the level-1 momentum generator remains a symmetry of the above Feynman integrals even if the conformal condition for the propagator powers in \eqref{eq:propandcond} does not hold, i.e.\ if there is no level-0 (dual) conformal symmetry. This statement applies to the massless and massive version of the Yangian symmetry and implies powerful constraints \cite{Loebbert:2020hxk}. Translating the $x$-space level-one momentum $\gen{\widehat P}^\mu$ to the dual momentum variables defined via $p_j^\mu=x_j^\mu-x_{j+1}^\mu$, one finds a massive generalization of the special conformal generator $\gen{\bar K}_{j}^\mu$  in momentum space which forms part of the following set of operators obeying the conformal algebra relations:
\begin{align}
\gen{\bar J}^a =\sum_{j=1}^n \gen{\bar J}_j^a
\qquad \text{with}\qquad
\gen{\bar J}_j^a \in
\begin{cases}
&\gen{\bar D}_{j} = p_{j\nu} \partial^{\nu}_{p_j}+ \sfrac{m_j \partial_{m_j}+m_{j+1} \partial_{m_{j+1}}}{2} + \bar{\Delta}_j,
\nonumber\\
&\gen{\bar L}_{j}^{\mu\nu} =  p_j^\mu \partial_{p_j}^{\nu} -  p_j^\nu \partial_{p_j}^{\mu},
\nonumber\\
&\gen{\bar P}_{j}^\mu = p_j^\mu \, , 
\nonumber\\
&\gen{\bar K}_{j}^\mu 
= p_j^\mu \partial_{p_j}^2
-2\brk[s]2{p_{j\nu} \partial_{p_j}^\nu+\sfrac{m_j \partial_{m_j}+m_{j+1} \partial_{m_{j+1}}}{2} + \bar{\Delta}_j }\partial_{p_j}^{\mu}.
\end{cases}
\label{eq:momspacemassive}
\end{align}
Note that these generator densities feature a nearest-neighbor action on the masses of the external legs of the Feynman graph.
In the context of $\superN=4$ SYM theory it is a well known statement that ordinary and dual conformal symmetry of scattering amplitudes close into the Yangian algebra \cite{Drummond:2009fd}. Similar statements hold for the above Feynman integrals in the case of the ordinary (bosonic) conformal algebra \cite{Chicherin:2017cns,Loebbert:2020glj}. The constraints from $\gen{\widehat P}^\mu$ or equivalently $\gen{\bar K}^\mu$ can thus be exploited independently of the dual conformal symmetry for massless or massive integrals. Here one representation or the other may be more convenient depending on whether one works in $x$- or $p$-space. Solving the momentum space conformal Ward identities in the massless case has recently been subject of great interest, see e.g.\ \cite{Coriano:2013jba,Bzowski:2013sza} and follow-ups.

\paragraph{Dimensional Regularization.}

Since the above symmetries can be formulated in generic spacetime dimension $D$, one can also employ them in the context of dimensionally regulated integrals. An interesting example is the following family of Euclidean three-point integrals with half integer propagator powers $a_j$ in $D=3-2\epsilon$ dimensions:
\begin{equation}
I_3^D[a_1,a_2,a_3] :=\int \frac{\dd^D{x}_0 }{({x}_{01}^2)^{a_1}({x}_{02}^2)^{a_2}({x}_{03}^2)^{a_3}}.
\end{equation}
Here we assume that
$a_1+a_2+a_3\leq {D}/{2}.$
The motivation to study these integrals comes from gravitational physics. The
Post-Newtonian (PN)  expansion with velocities $v\ll c$ of the 3-body effective potential in general relativity requires these integrals as an input, e.g.\ in the following contribution (in red) to the 3PN potential \cite{Loebbert:2020aos}:
\begin{align}
    &S^\text{3PN}=\dots+\sum_{\atopfrac{j\neq i}{k\neq i, j}}
    \frac{G^2 m_i m_j m_k}{ 4\pi } 
    \bigg\{ 
    \Big[ 
         ( 6 \mbf{v}_i^2 \!+\! \mbf{v}_k^2 \!-\! 8 \mbf{v}_i \!\cdot\! \mbf{v}_j ) 
            (\mbf{v}_{ki} \!\cdot\! \partial_{x_i})(\mbf{v}_{kj} \!\cdot\! \partial_{x_j})
            \nonumber\\[-4mm]
       &\hspace{4cm}
       + ( 8 \mbf{v}_{ik}^2 \!-\! 4 \mbf{v}_k^2 ) 
            (\mbf{v}_{ji} \!\cdot\! \partial_{x_i}) 
            (\mbf{v}_{ij} \!\cdot\! \partial_{x_j})
    \Big] \highlight{I_3[\half, \half, \half]}
    \nonumber \\ & 
   \hspace{2cm} + \left( \mbf{v}_{k} \!\cdot\! \partial_{x_k} \right)^2
    \Big[	
           (\mbf{v}_{ki} \!\cdot\! \partial_{x_i}) (\mbf{v}_{kj} \!\cdot\! \partial_{x_j})
        \!+\! 2(\mbf{v}_{ik} \!\cdot\! \partial_{x_k}) (\mbf{v}_{ij} \!\cdot\! \partial_{x_j})
        \nonumber\\
       & \hspace{2cm}\!+\! 4(\mbf{v}_{ji} \!\cdot\! \partial_{x_i}) (\mbf{v}_{ij} \!\cdot\! \partial_{x_j})
        \!+\! 8(\mbf{v}_{jk} \!\cdot\! \partial_{x_k}) (\mbf{v}_{kj} \!\cdot\! \partial_{x_j})
    \Big] \highlight{I_3[\half, \half,\! -\half]}
    \bigg\}. 
    \label{eq:3bodycontr}
\end{align}
To bootstrap these integrals one can employ the Yangian level-1 momentum generator, which yields two 
second order PDEs in the variables $r_{jk} = \abs{x_{j}-x_{k}}$ (see \cite{Loebbert:2020aos} for explicit expressions):
\begin{equation}
D_j I_3=0, \qquad j=1,2.
\end{equation}
These differential equations are solved by the ansatz
\begin{equation*}
    \mu^{-2\epsilon}I_3^{3-2\epsilon}[a_1,a_2,a_3] =  \frac{A}{2\epsilon} + B + C \log\brk1{\sfrac{r_{12}+r_{13}+r_{23}}{\mu}}+\order{\epsilon},
\end{equation*}
with a scale $\mu$ and given polynomials $A,B,C$, which is inspired by the leading integral in this family as presented in \cite{Ohta:1973je}.
Plugging the resulting integrals into the effective potential via \eqref{eq:3bodycontr}, this leads to new $v^{2n}G^2$ contributions to the Post-Newtonian expansion of the 3-body effective potential.
The same approach can be applied to higher PN integrals $I_3^{3-2\epsilon}[a_1,a_2,a_3]$ in this family (cf.\ \cite{Loebbert:2020aos}), which can then be inserted into the higher order analogues of \eqref{eq:3bodycontr}.


\paragraph{Soft and Collinear Anomalies.}

Notably, the above Yangian or momentum space conformal symmetry does not commute with coincidence or lightlike limits of the external kinematic variables $x_j^\mu$ or their differences, respectively. These correspond to soft or on-shell limits in the dual momentum variables $p_j^\mu$:
\begin{align}
&\text{Coincidence/soft limit:} &&x_j^\mu\to x_{j+1}^\mu && \leftrightarrow && p_j^\mu\to0,
\\
&\text{Lightlike/on-shell limit:} &&x_{j,j+1}^2 \to 0 && \leftrightarrow && p_j^2 \to0.
\end{align}
These situations lead to anomalies even for finite integrals, see \cite{Corcoran:2021gda} for examples of coincidence limits and \cite{Chicherin:2017bxc,Chicherin:2022nqq}  for the on-shell case.

\section{Yangian Symmetry and Calabi--Yau Geometry}
\label{sec:CY}

In order to make progress on understanding fishnet integrals at higher loop orders, we consider Feynman graphs in $D=2$ spacetime dimensions, see \cite{Duhr:2022pch} for more details. For square fishnets with propagators $(x^{-2})^{a_j}$ to obey the conformal condition $ \sum_{j=1}^4a_j=D$ we set $a_j=1/2$ for all propagators.
The respective integrals represent correlators in the $D$-dimensional version of the square fishnet theory defined by the Lagrangian in \eqref{eq:DFishLagr} \cite{Kazakov:2018qbr}.
Similar to the conformal Lie algebra in two dimensions, the conformal Yangian splits into a holomorphic and an anti-holomorphic part: $Y[\alg{sl}(2,\mathbb{R})]\oplus \overline{Y[\alg{sl}(2,\mathbb{R})]}$.
We thus expect the following double copy form for the respective fishnet integrals:
\begin{equation}
\phi(z,\bar z)=\vec{\Pi}^\dagger(\bar z) \cdot  \Sigma \cdot \vec{\Pi}(z).
\label{eq:pisigpi}
\end{equation}
Here $\vec{\Pi}(z)$ denotes a vector of solutions to the Yangian invariance equations for $Y[\alg{sl}(2,\mathbb{R})]$ with $z$ indicating the conformal cross ratios and $\bar z$ their complex conjugates. The symbol $\Sigma$ represents a constant matrix which defines the precise linear combination of the Yangian invariants representing the Feynman integral. 

\paragraph{Example: 2D Cross (or Box) Integral.}

Consider the simplest example of the 2D cross integral which, using conformal symmetry, can be written in the form
\begin{equation}
I_4^{2D}=
\includegraphicsbox[scale=1.2]{FigFourPointVertexNox}
=\frac{1}{\abs{x_{12}}\abs{x_{34}}}\phi(z,\bar z).
\end{equation}
The cross integral depends on a single conformal variable $z$ (and its conjugate $\bar z$). Imposing invariance of the holomorphic part of the integral under $Y[\alg{sl}(2,\mathbb{R})]$, the respective Yangian differential equation takes the form
\begin{equation}
\brk*{1+4(2z-1)\partial_z+4z(z-1)\partial_z^2}\vec{\Pi}(z)=0.
\label{eq:Yangdiffbox}
\end{equation}
One finds two solutions $K(z)$ and $K(1-z)$ to this equation, which indeed combine into a double copy of Yangian invariants for the 2D cross integral \cite{Derkachov:2018rot,Corcoran:2021gda}:
\begin{equation}
\phi(z,\bar z)
=
\frac{4}{\pi}i\begin{pmatrix}
K(z)&iK(1-z)
\end{pmatrix}
\cdot 
\begin{pmatrix}
0&1
\\
-1&0
\end{pmatrix}
\cdot
\begin{pmatrix}
K(\bar z)
\\
-iK(1-\bar z)
\end{pmatrix}.
\end{equation}
The two solutions $K(z)$ and $K(1-z)$ of the Yangian equation are identified with the complete elliptic integral of the first kind evaluated at different arguments:
\begin{equation}
K(z)
=\int_0^1 \frac{dt}{\sqrt{(1-t^2)(1-zt^2)}}.
\end{equation}
This integral has an interesting relation to geometry via the elliptic curve defined by the polynomial in the denominator of the integrand:
\vspace{-4mm}
\begin{equation}
y^2=(1-t^2)(1-z t^2),
\hspace{1.3cm}
\includegraphicsbox[scale=1.5]{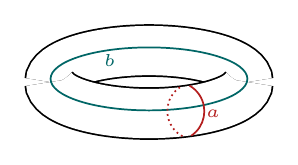}
\vspace{-2mm}
\end{equation}
The elliptic curve is isomorphic to a torus with two distinguished cycles denoted by $a$ and $b$.
The two period integrals associated to these two cycles are precisely the two Yangian invariants $K(z)$ and $K(1-z)$ with \eqref{eq:Yangdiffbox} representing the so-called Picard--Fuchs differential equation associated with the elliptic curve. 
\medskip

For higher loop fishnet integrals in 2D it turns out that the above double copy pattern continues to hold. Also these integrals can be written in the form \eqref{eq:pisigpi} with a vector of Yangian invariant functions $\vec{\Pi}$ of the conformal cross ratios \cite{Duhr:2022pch}.%
\footnote{To obtain the complete Picard--Fuchs ideal of differential operators one has to take the permutation symmetries of the integrals into account, see also the example of the 4D cross integral given in \Secref{sec:intforF} where these were used in a different way.}
The geometric interpretation of these Yangian invariants generalizes to a family of Calabi--Yau's with the above elliptic curve representing a Calabi--Yau $1$-fold. We note that a Calabi--Yau $\ell$-fold is defined as an $\ell$-dimensional complex K\"ahler manifold with vanishing first Chern class. Here the last condition relates to the conformal nature of the fishnet integrals defined in terms of four-point vertices only. As for the elliptic curve, also the higher Calabi--Yau's posses an associated set of period integrals which are annihilated by a Picard--Fuchs ideal of differential operators. The conformal cross ratios of the fishnet integrals correspond to the Calabi--Yau moduli. At 3 loops for instance, the respective 2D integral depends on 5 cross ratios and corresponds to a Calabi--Yau 3-fold with a 12-dimensional vector $\vec{\Pi}$ of Yangian-invariant period integrals. The conjecture of \cite{Duhr:2022pch} suggests that for generic 2D fishnet integrals the Picard--Fuchs ideal is obtained by combining Yangian and permutation symmetries of the respective Feynman graph. This represents a curious new relation between integrability and Calabi--Yau geometries!

Notably, the geometric interpretation of fishnet integrals goes even further \cite{Duhr:2022pch}. The fishnet expression of \eqref{eq:pisigpi} corresponds to the so-called quantum volume of a particular Calabi--Yau geometry, which is obtained from the original fishnet Calabi--Yau via the celebrated mirror symmetry. This finding extends the volume interpretation of Feynman integrals, which was well known at one loop order (see e.g.\ \cite{Bourjaily:2019exo}), to higher loops --- at least in two spacetime dimensions. Another notable fact is that the conformal fishnet function $\phi$ computes the K\"ahler potential $V$ of the Calabi--Yau via the relation $\phi=\exp(-V)$. More details on this elaborate construction are given in \cite{Duhr:2022pch}.


\section{Conclusion}

It is fascinating that integrable structures appear for rather broad classes of Feynman integrals, which constitute the building blocks of generic quantum field theories. The two-dimensional nature of integrability is satisfied with the ordering of the external legs of planar Feynman graphs. Here we have focussed on the Yangian invariance of Feynman integrals, which provides systems of partial differential equations originating in a spacetime symmetry. 
We note that integrability in the context of fishnet integrals can be considered from various different angles and already in 1980 A.B.~Zamolodchikov investigated integrable structures for these integrals in the context of statistical vertex models \cite{Zamolodchikov:1980mb}. More recent applications include the extraction of Feynman integrals from elements of the fishnet dilatation operator \cite{Caetano:2016ydc}, the Basso--Dixon correlators of \cite{Basso:2017jwq, Derkachov:2018rot}, or the interesting number-theoretic relation to $Q$-functions of \cite{Gurdogan:2020ppd}.

There are numerous intriguing directions to further explore the Yangian symmetry of Feynman integrals. In particular, the connection to Calabi--Yau geometries promises interesting novel insights and new connections to mathematics.

Another curious appearance of integrability in the context of higher dimensional conformal field theory is the interpretation of conformal blocks as eigenfunctions of integrable Calogero--Sutherland or Gaudin Hamiltonians \cite{Isachenkov:2016gim,Buric:2020dyz}. It would be fascinating to relate these structures to the Yangian symmetry of conformal correlation functions discussed above.

Going beyond the realm of flat space, there are curious new relations to correlation functions in Anti-de-Sitter space.
In \cite{Rigatos:2022eos} it was noted that Witten contact diagrams are identical with one-loop Feynman integrals featuring Yangian symmetry \cite{Loebbert:2020hxk}. If this connection generalizes to other classes of Witten diagrams, this may open a new playground for applications of integrability in AdS spaces.


\section*{Acknowledgements}
First, I would like to thank the organizers of the \emph{34th International Colloquium on Group Theoretical Methods in Physics} in Strasbourg for the invitation to this stimulating interdisciplinary conference.
I am very grateful to the collaborators who worked with me on the interesting topic discussed here, that is
Dima Chicherin,
Luke Corcoran,
Claude Duhr,
Volodya Kazakov, 
Albrecht Klemm,
Julian Miczajka,
Dennis M\"uller, 
Hagen M\"unkler, 
Christoph Nega,
Jan Plefka,
Franziska Porkert,
Canxin Shi, 
Matthias Staudacher,
Tianheng Wang and
De-liang Zhong.


The work of FL is funded by the Deutsche Forschungsgemeinschaft (DFG, German Research Foundation)-Projektnummer 363895012, and by funds of the Klaus Tschira Foundation gGmbH.

\bibliography{ProceedingsGroup34}

\begin{thebibliography}{10}
\providecommand{\url}[1]{\texttt{#1}}
\providecommand{\urlprefix}{URL }
\expandafter\ifx\csname urlstyle\endcsname\relax
  \providecommand{\doi}[1]{doi:\discretionary{}{}{}#1}\else
  \providecommand{\doi}{doi:\discretionary{}{}{}\begingroup
  \urlstyle{rm}\Url}\fi
\providecommand{\eprint}[2][]{\url{#2}}

\bibitem{Lipatov:1993yb}
L.~N. Lipatov,
\newblock \emph{{Asymptotic behavior of multicolor QCD at high energies in
  connection with exactly solvable spin models}},
\newblock JETP Lett. \textbf{59}, 596 (1994),
\newblock \eprint{hep-th/9311037}.

\bibitem{Beisert:2010jr}
N.~Beisert \emph{et~al.},
\newblock \emph{{Review of AdS/CFT Integrability: An Overview}},
\newblock Lett. Math. Phys. \textbf{99}, 3 (2012),
\newblock \doi{10.1007/s11005-011-0529-2},
\newblock \eprint{1012.3982}.

\bibitem{Gurdogan:2015csr}
O.~G\"urdo\u{g}an and V.~Kazakov,
\newblock \emph{{New Integrable 4D Quantum Field Theories from Strongly
  Deformed Planar $\mathcal N = $ 4 Supersymmetric Yang-Mills Theory}},
\newblock Phys. Rev. Lett. \textbf{117}(20), 201602 (2016),
\newblock \doi{10.1103/PhysRevLett.117.201602},
\newblock [Addendum: Phys.Rev.Lett. 117, 259903 (2016)],
\newblock \eprint{1512.06704}.

\bibitem{Caetano:2016ydc}
J.~a. Caetano, O.~G\"urdo\u{g}an and V.~Kazakov,
\newblock \emph{{Chiral limit of $ \mathcal{N} $ = 4 SYM and ABJM and
  integrable Feynman graphs}},
\newblock JHEP \textbf{03}, 077 (2018),
\newblock \doi{10.1007/JHEP03(2018)077},
\newblock \eprint{1612.05895}.

\bibitem{Chicherin:2017cns}
D.~Chicherin, V.~Kazakov, F.~Loebbert, D.~M\"uller and D.-l. Zhong,
\newblock \emph{{Yangian Symmetry for Bi-Scalar Loop Amplitudes}},
\newblock JHEP \textbf{05}, 003 (2018),
\newblock \doi{10.1007/JHEP05(2018)003},
\newblock \eprint{1704.01967}.

\bibitem{Bourjaily:2022bwx}
J.~L. Bourjaily \emph{et~al.},
\newblock \emph{{Functions Beyond Multiple Polylogarithms for Precision
  Collider Physics}},
\newblock In \emph{{2022 Snowmass Summer Study}} (2022), \eprint{2203.07088}.

\bibitem{sutherland2004beautiful}
B.~Sutherland,
\newblock \emph{Beautiful Models: 70 Years of Exactly Solved Quantum Many-body
  Problems},
\newblock World Scientific (2004).

\bibitem{Caux:2010by}
J.-S. Caux and J.~Mossel,
\newblock \emph{{Remarks on the notion of quantum integrability}},
\newblock J. Stat. Mech. \textbf{1102}, P02023 (2011),
\newblock \doi{10.1088/1742-5468/2011/02/P02023},
\newblock \eprint{1012.3587}.

\bibitem{Coleman:1967ad}
S.~R. Coleman and J.~Mandula,
\newblock \emph{{All Possible Symmetries of the S Matrix}},
\newblock Phys. Rev. \textbf{159}, 1251 (1967),
\newblock \doi{10.1103/PhysRev.159.1251}.

\bibitem{Loebbert:2018lxk}
F.~Loebbert and A.~Spiering,
\newblock \emph{{Nonlocal Symmetries and Factorized Scattering}},
\newblock J. Phys. A \textbf{51}(48), 485202 (2018),
\newblock \doi{10.1088/1751-8121/aae7ff},
\newblock \eprint{1805.11993}.

\bibitem{Drinfeld:1985rx}
V.~G. Drinfeld,
\newblock \emph{{Hopf algebras and the quantum Yang-Baxter equation}},
\newblock Sov. Math. Dokl. \textbf{32}, 254 (1985).

\bibitem{Bernard:1992ya}
D.~Bernard,
\newblock \emph{{An Introduction to Yangian Symmetries}},
\newblock Int. J. Mod. Phys. B \textbf{7}, 3517 (1993),
\newblock \doi{10.1142/S0217979293003371},
\newblock \eprint{hep-th/9211133}.

\bibitem{MacKay:2004tc}
N.~J. MacKay,
\newblock \emph{{Introduction to Yangian symmetry in integrable field theory}},
\newblock Int. J. Mod. Phys. A \textbf{20}, 7189 (2005),
\newblock \doi{10.1142/S0217751X05022317},
\newblock \eprint{hep-th/0409183}.

\bibitem{Loebbert:2016cdm}
F.~Loebbert,
\newblock \emph{{Lectures on Yangian Symmetry}},
\newblock J. Phys. A \textbf{49}(32), 323002 (2016),
\newblock \doi{10.1088/1751-8113/49/32/323002},
\newblock \eprint{1606.02947}.

\bibitem{Miczajka:2022jjv}
J.~Miczajka,
\newblock \emph{{The Yangian Bootstrap for Massive Feynman Diagrams}},
\newblock Ph.D. thesis, Humboldt U., Berlin,
\newblock \doi{10.18452/23855} (2022).

\bibitem{Dokmetzoglou:2022mfd}
N.~Dokmetzoglou and L.~Dolan,
\newblock \emph{{Properties of the Conformal Yangian in Scalar and Gauge Field
  Theories}}  (2022),
\newblock \eprint{2207.14806}.

\bibitem{Torrielli:2010kq}
A.~Torrielli,
\newblock \emph{{Review of AdS/CFT Integrability, Chapter VI.2: Yangian
  Algebra}},
\newblock Lett. Math. Phys. \textbf{99}, 547 (2012),
\newblock \doi{10.1007/s11005-011-0491-z},
\newblock \eprint{1012.4005}.

\bibitem{Dolan:2003uh}
L.~Dolan, C.~R. Nappi and E.~Witten,
\newblock \emph{{A Relation between approaches to integrability in
  superconformal Yang-Mills theory}},
\newblock JHEP \textbf{10}, 017 (2003),
\newblock \doi{10.1088/1126-6708/2003/10/017},
\newblock \eprint{hep-th/0308089}.

\bibitem{Sieg:2016vap}
C.~Sieg and M.~Wilhelm,
\newblock \emph{{On a CFT limit of planar $\gamma_i$-deformed $\mathcal{N}=4$
  SYM theory}},
\newblock Phys. Lett. B \textbf{756}, 118 (2016),
\newblock \doi{10.1016/j.physletb.2016.03.004},
\newblock \eprint{1602.05817}.

\bibitem{Grabner:2017pgm}
D.~Grabner, N.~Gromov, V.~Kazakov and G.~Korchemsky,
\newblock \emph{{Strongly $\gamma$-Deformed $\mathcal{N}=4$ Supersymmetric
  Yang-Mills Theory as an Integrable Conformal Field Theory}},
\newblock Phys. Rev. Lett. \textbf{120}(11), 111601 (2018),
\newblock \doi{10.1103/PhysRevLett.120.111601},
\newblock \eprint{1711.04786}.

\bibitem{Loebbert:2020hxk}
F.~Loebbert, J.~Miczajka, D.~M\"uller and H.~M\"unkler,
\newblock \emph{{Massive Conformal Symmetry and Integrability for Feynman
  Integrals}},
\newblock Phys. Rev. Lett. \textbf{125}(9), 091602 (2020),
\newblock \doi{10.1103/PhysRevLett.125.091602},
\newblock \eprint{2005.01735}.

\bibitem{Loebbert:2019vcj}
F.~Loebbert, D.~M\"uller and H.~M\"unkler,
\newblock \emph{{Yangian Bootstrap for Conformal Feynman Integrals}},
\newblock Phys. Rev. D \textbf{101}(6), 066006 (2020),
\newblock \doi{10.1103/PhysRevD.101.066006},
\newblock \eprint{1912.05561}.

\bibitem{Usyukina:1992jd}
N.~I. Usyukina and A.~I. Davydychev,
\newblock \emph{{An Approach to the evaluation of three and four point ladder
  diagrams}},
\newblock Phys. Lett. B \textbf{298}, 363 (1993),
\newblock \doi{10.1016/0370-2693(93)91834-A}.

\bibitem{Corcoran:2020akn}
L.~Corcoran and M.~Staudacher,
\newblock \emph{{The dual conformal box integral in Minkowski space}},
\newblock Nucl. Phys. B \textbf{964}, 115310 (2021),
\newblock \doi{10.1016/j.nuclphysb.2021.115310},
\newblock \eprint{2006.11292}.

\bibitem{Corcoran:2020epz}
L.~Corcoran, F.~Loebbert, J.~Miczajka and M.~Staudacher,
\newblock \emph{{Minkowski Box from Yangian Bootstrap}},
\newblock JHEP \textbf{04}, 160 (2021),
\newblock \doi{10.1007/JHEP04(2021)160},
\newblock \eprint{2012.07852}.

\bibitem{Chicherin:2017frs}
D.~Chicherin, V.~Kazakov, F.~Loebbert, D.~M\"uller and D.-l. Zhong,
\newblock \emph{{Yangian Symmetry for Fishnet Feynman Graphs}},
\newblock Phys. Rev. D \textbf{96}(12), 121901 (2017),
\newblock \doi{10.1103/PhysRevD.96.121901},
\newblock \eprint{1708.00007}.

\bibitem{Mamroud:2017uyz}
O.~Mamroud and G.~Torrents,
\newblock \emph{{RG stability of integrable fishnet models}},
\newblock JHEP \textbf{06}, 012 (2017),
\newblock \doi{10.1007/JHEP06(2017)012},
\newblock \eprint{1703.04152}.

\bibitem{Kazakov:2018qbr}
V.~Kazakov and E.~Olivucci,
\newblock \emph{{Biscalar Integrable Conformal Field Theories in Any
  Dimension}},
\newblock Phys. Rev. Lett. \textbf{121}(13), 131601 (2018),
\newblock \doi{10.1103/PhysRevLett.121.131601},
\newblock \eprint{1801.09844}.

\bibitem{Alday:2009zm}
L.~F. Alday, J.~M. Henn, J.~Plefka and T.~Schuster,
\newblock \emph{{Scattering into the fifth dimension of N=4 super Yang-Mills}},
\newblock JHEP \textbf{01}, 077 (2010),
\newblock \doi{10.1007/JHEP01(2010)077},
\newblock \eprint{0908.0684}.

\bibitem{Loebbert:2020tje}
F.~Loebbert and J.~Miczajka,
\newblock \emph{{Massive Fishnets}},
\newblock JHEP \textbf{12}, 197 (2020),
\newblock \doi{10.1007/JHEP12(2020)197},
\newblock \eprint{2008.11739}.

\bibitem{Loebbert:2020glj}
F.~Loebbert, J.~Miczajka, D.~M\"uller and H.~M\"unkler,
\newblock \emph{{Yangian Bootstrap for Massive Feynman Integrals}},
\newblock SciPost Phys. \textbf{11}, 010 (2021),
\newblock \doi{10.21468/SciPostPhys.11.1.010},
\newblock \eprint{2010.08552}.

\bibitem{Karananas:2019fox}
G.~K. Karananas, V.~Kazakov and M.~Shaposhnikov,
\newblock \emph{{Spontaneous Conformal Symmetry Breaking in Fishnet CFT}},
\newblock Phys. Lett. B \textbf{811}, 135922 (2020),
\newblock \doi{10.1016/j.physletb.2020.135922},
\newblock \eprint{1908.04302}.

\bibitem{Drummond:2009fd}
J.~M. Drummond, J.~M. Henn and J.~Plefka,
\newblock \emph{{Yangian symmetry of scattering amplitudes in N=4 super
  Yang-Mills theory}},
\newblock JHEP \textbf{05}, 046 (2009),
\newblock \doi{10.1088/1126-6708/2009/05/046},
\newblock \eprint{0902.2987}.

\bibitem{Coriano:2013jba}
C.~Coriano, L.~Delle~Rose, E.~Mottola and M.~Serino,
\newblock \emph{{Solving the Conformal Constraints for Scalar Operators in
  Momentum Space and the Evaluation of Feynman's Master Integrals}},
\newblock JHEP \textbf{07}, 011 (2013),
\newblock \doi{10.1007/JHEP07(2013)011},
\newblock \eprint{1304.6944}.

\bibitem{Bzowski:2013sza}
A.~Bzowski, P.~McFadden and K.~Skenderis,
\newblock \emph{{Implications of conformal invariance in momentum space}},
\newblock JHEP \textbf{03}, 111 (2014),
\newblock \doi{10.1007/JHEP03(2014)111},
\newblock \eprint{1304.7760}.

\bibitem{Loebbert:2020aos}
F.~Loebbert, J.~Plefka, C.~Shi and T.~Wang,
\newblock \emph{{Three-body effective potential in general relativity at second
  post-Minkowskian order and resulting post-Newtonian contributions}},
\newblock Phys. Rev. D \textbf{103}(6), 064010 (2021),
\newblock \doi{10.1103/PhysRevD.103.064010},
\newblock \eprint{2012.14224}.

\bibitem{Ohta:1973je}
T.~Ohta, H.~Okamura, T.~Kimura and K.~Hiida,
\newblock \emph{{Physically acceptable solution of einstein's equation for
  many-body system}},
\newblock Prog. Theor. Phys. \textbf{50}, 492 (1973),
\newblock \doi{10.1143/PTP.50.492}.

\bibitem{Corcoran:2021gda}
L.~Corcoran, F.~Loebbert and J.~Miczajka,
\newblock \emph{{Yangian Ward identities for fishnet four-point integrals}},
\newblock JHEP \textbf{04}, 131 (2022),
\newblock \doi{10.1007/JHEP04(2022)131},
\newblock \eprint{2112.06928}.

\bibitem{Chicherin:2017bxc}
D.~Chicherin and E.~Sokatchev,
\newblock \emph{{Conformal anomaly of generalized form factors and finite loop
  integrals}},
\newblock JHEP \textbf{04}, 082 (2018),
\newblock \doi{10.1007/JHEP04(2018)082},
\newblock \eprint{1709.03511}.

\bibitem{Chicherin:2022nqq}
D.~Chicherin and G.~P. Korchemsky,
\newblock \emph{{Chapter 9: Integrability of amplitudes in fishnet theories}},
\newblock J. Phys. A \textbf{55}(44), 443010 (2022),
\newblock \doi{10.1088/1751-8121/ac8c72},
\newblock \eprint{2203.13020}.

\bibitem{Duhr:2022pch}
C.~Duhr, A.~Klemm, F.~Loebbert, C.~Nega and F.~Porkert,
\newblock \emph{{Yangian-Invariant Fishnet Integrals in Two Dimensions as
  Volumes of Calabi-Yau Varieties}},
\newblock Phys. Rev. Lett. \textbf{130}(4), 041602 (2023),
\newblock \doi{10.1103/PhysRevLett.130.041602},
\newblock \eprint{2209.05291}.

\bibitem{Derkachov:2018rot}
S.~Derkachov, V.~Kazakov and E.~Olivucci,
\newblock \emph{{Basso-Dixon Correlators in Two-Dimensional Fishnet CFT}},
\newblock JHEP \textbf{04}, 032 (2019),
\newblock \doi{10.1007/JHEP04(2019)032},
\newblock \eprint{1811.10623}.

\bibitem{Bourjaily:2019exo}
J.~L. Bourjaily, E.~Gardi, A.~J. McLeod and C.~Vergu,
\newblock \emph{{All-mass $n$-gon integrals in $n$ dimensions}},
\newblock JHEP \textbf{08}(08), 029 (2020),
\newblock \doi{10.1007/JHEP08(2020)029},
\newblock \eprint{1912.11067}.

\bibitem{Zamolodchikov:1980mb}
A.~B. Zamolodchikov,
\newblock \emph{{'Fishing-net' Diagrams as a Completely Integrable System}},
\newblock Phys. Lett. B \textbf{97}, 63 (1980),
\newblock \doi{10.1016/0370-2693(80)90547-X}.

\bibitem{Basso:2017jwq}
B.~Basso and L.~J. Dixon,
\newblock \emph{{Gluing Ladder Feynman Diagrams into Fishnets}},
\newblock Phys. Rev. Lett. \textbf{119}(7), 071601 (2017),
\newblock \doi{10.1103/PhysRevLett.119.071601},
\newblock \eprint{1705.03545}.

\bibitem{Gurdogan:2020ppd}
O.~G\"urdo\u{g}an,
\newblock \emph{{From integrability to the Galois coaction on Feynman
  periods}},
\newblock Phys. Rev. D \textbf{103}(8), L081703 (2021),
\newblock \doi{10.1103/PhysRevD.103.L081703},
\newblock \eprint{2011.04781}.

\bibitem{Isachenkov:2016gim}
M.~Isachenkov and V.~Schomerus,
\newblock \emph{{Superintegrability of $d$-dimensional Conformal Blocks}},
\newblock Phys. Rev. Lett. \textbf{117}(7), 071602 (2016),
\newblock \doi{10.1103/PhysRevLett.117.071602},
\newblock \eprint{1602.01858}.

\bibitem{Buric:2020dyz}
I.~Buric, S.~Lacroix, J.~A. Mann, L.~Quintavalle and V.~Schomerus,
\newblock \emph{{From Gaudin Integrable Models to $d$-dimensional Multipoint
  Conformal Blocks}},
\newblock Phys. Rev. Lett. \textbf{126}(2), 021602 (2021),
\newblock \doi{10.1103/PhysRevLett.126.021602},
\newblock \eprint{2009.11882}.

\bibitem{Rigatos:2022eos}
K.~C. Rigatos and X.~Zhou,
\newblock \emph{{Yangian Symmetry in Holographic Correlators}},
\newblock Phys. Rev. Lett. \textbf{129}(10), 101601 (2022),
\newblock \doi{10.1103/PhysRevLett.129.101601},
\newblock \eprint{2206.07924}.

\end{thebibliography}

\nolinenumbers

\end{document}